\documentclass{vgtc} 
\usepackage{mathptmx}                  
\usepackage{appendix}
\usepackage{amsmath,amsfonts}
\usepackage{algorithmic}
\usepackage{algorithm}
\usepackage{array}
\usepackage[caption=false,font=normalsize,labelfont=sf,textfont=sf]{subfig}
\usepackage{textcomp}
\usepackage{stfloats}
\usepackage{url}
\usepackage{verbatim}
\usepackage{graphicx}
\usepackage{cite}
\usepackage{xcolor}
\usepackage{url}

\onlineid{1117}

\vgtccategory{Research}

\title{Exploring Gaze Dynamics in VR Film Education: Gender, Avatar, and the Shift Between Male and Female Perspectives}

\author{Zheng Wei\thanks{e-mail: zwei302@connect.ust.hk}\\ %
        \scriptsize Hong Kong University of \\ %
        \scriptsize Science and Technology %
\and Jia Sun\thanks{e-mail: jsun666@connect.hkust-gz.edu.cn}\\ \scriptsize Hong Kong University of \\ %
        \scriptsize Science and Technology (Guangzhou) %
\and Junxiang Liao\thanks{e-mail: junxiang.liao@connect.polyu.hk}\\ %
     \scriptsize 
The Hong Kong Polytechnic University \\ %
\and Lik-Hang Lee\thanks{e-mail: lik-hang.lee@polyu.edu.hk}\\ %
     \scriptsize The Hong Kong Polytechnic University %
\and Pan Hui\thanks{e-mail: panhui@hkust-gz.edu.cn}\\%
     \scriptsize Hong Kong University of \\ \scriptsize Science and Technology (Guangzhou) %
\and Huamin Qu\thanks{e-mail: huamin@cse.ust.hk}\\%
     \scriptsize  Hong Kong University of \\ \scriptsize Science and Technology %
\and Wai Tong\thanks{e-mail: wtong@tamu.edu}\\%
     \scriptsize  Texas A\&M University%
\and Xian Xu\thanks{e-mail: xianxu0523@gmail.com}\\ %
     \scriptsize  Lingnan University  \thanks{Xian Xu and Wai Tong are corresponding authors.}%
     }

\teaser{
  \centering
  \includegraphics[width=\linewidth]{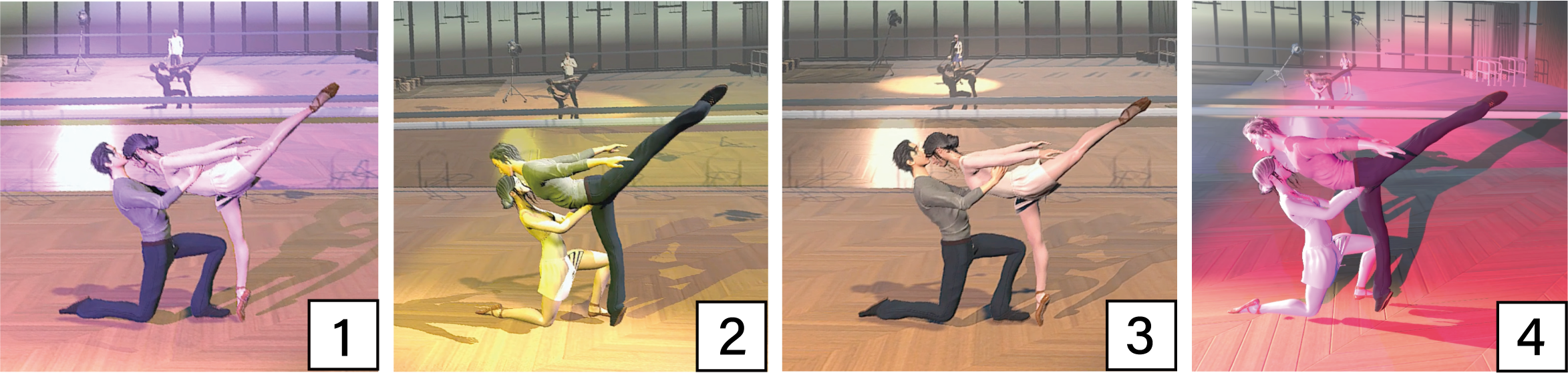}
  \caption{The figure shows four of eight VR film production conditions: (1) male user with male avatar and male-led narrative; (2) male user with male avatar and female-led narrative; (3) female user with female avatar and male-led narrative; (4) female user with female avatar and female-led narrative.}
  \label{fig:teaser}
}

\abstract{In virtual reality (VR) education, especially in creative fields like film production, avatar design and narrative style extend beyond appearance and aesthetics. This study explores how the interaction between avatar gender, the dominant narrative actor's gender, and the learner's gender influences film production learning in VR, focusing on gaze dynamics and gender perspectives. Using a 2$\times$2$\times$2 experimental design, 48 participants operated avatars of different genders and interacted with male or female-dominant narratives. The results show that the consistency between the avatar and gender affects presence, and learners' control over the avatar is also influenced by gender matching. Learners using avatars of the opposite gender reported stronger control, suggesting gender incongruity prompted more focus on the avatar. Additionally, female participants with female avatars were more likely to adopt a ``female gaze,'' favoring soft lighting and emotional shots, while male participants with male avatars were more likely to adopt a ``male gaze,'' choosing dynamic shots and high contrast. When male participants used female avatars, they favored ``female gaze,'' while female participants with male avatars focused on ``male gaze''. These findings advance our understanding of how avatar design and narrative style in VR-based education influence creativity and the cultivation of gender perspectives, and they offer insights for developing more inclusive and diverse VR teaching tools going forward.}

\graphicspath{{figures/}{pictures/}{images/}{./}}

\begin{document}


\firstsection{Introduction}

\maketitle
In recent years, virtual reality (VR) technology has gradually become an important tool in the field of education, especially in film production \cite{wei2023feeling,wei2025illuminating}. In film production, complex tasks such as cinematography and lighting require not only technical expertise but also a strong sense of narrative creativity \cite{wei2024multi}. VR technology provides students with a controllable interactive environment, allowing them to immerse themselves in exploring different camera angles, lighting setups, and other cinematic elements \cite{xu2024transforming}. This immersive experience offers an extremely effective platform for skill development, which is often difficult to achieve in traditional classroom settings {\cite{wei2024hearing,fitrianto2024role,quintana2015pedagogical,di2020immersive}}. However, despite the widely recognized advantages of VR technology in education, research on how specific aspects of virtual environments, particularly avatar design and narrative context, impact students' learning experiences remains very limited. Especially in the creative task of film production, how VR can simultaneously promote both the improvement of technical skills and the cultivation of creative expression remains a critical challenge that needs to be addressed.

Self-presence (the feeling of ``being there'') plays a crucial role in how users experience VR \cite{berkman2024presence}. Previous research has shown that avatars that resemble the user's appearance (such as body type, skin tone, and facial features) significantly enhance their sense of immersion and connection with the virtual world \cite{kim2024self,do2024stepping,freeman2021body}. This sense of self-presence is particularly important in film education as it directly influences students' ability to make decisions and creative choices within virtual environments {\cite{wei2025illuminating,wei2024multi}}. When users think their avatars authentically represent themselves, their engagement and emotional investment in the task tend to be higher \cite{suh2011if}, especially when the task involves complex creative expression, such as film production. In addition to avatar design, the dynamics of gendered avatars and gendered narratives also significantly affect the user experience \cite{zhang2024gender}. The \textit{Proteus Effect} suggests that users' behavior and self-perception are influenced by the characteristics of their avatars, particularly gender characteristics \cite{yee2007proteus,yee2009proteus,bailenson2009virtual}. When users interact with avatars of different genders, they may unconsciously adopt behaviors associated with that gender, which in turn affects their attitudes and behavior within the virtual environment \cite{sherrick2014role}. These gender dynamics are especially important in educational settings, particularly in creative disciplines such as film production, where gender perspectives and gaze dynamics play a central role in storytelling \cite{yu2024unravelling,mistri2024evolution,newman1990situation,gibson2011disciplining}.

Despite the increasing research on avatar design and the \textit{Proteus Effect} \cite{ratan2020avatar}, there is still a lack of in-depth understanding of how these factors interact in narrative-driven VR environments, particularly in the context of film production. This study aims to explore the interaction between gendered avatars, narrative-driven content, and user self-presentation, focusing on how avatar gender and gender dynamics in narratives influence students' self-presence, cognitive engagement, and decision-making. Specifically, the study investigates how gender congruence between learners and avatars and the gender dynamics in narratives influence users' creative choices and their understanding of gender perspectives in film production. The following research questions guide this study: 
\textbf{RQ1}: How does the gender of participants and their avatars affect participants' self-perception, sense of presence, and immersion? \textbf{RQ2}: How do the gender of the dominant narrative role (male-driven vs. female-driven narrative) and the participant's gender and avatar gender interact to influence their creative decisions (e.g., camera shots, lighting, emotional emphasis)? \textbf{RQ3}: How do gendered avatars and narrative-driven content influence participants' understanding of gender perspectives in film production, particularly in terms of gendered gaze and power dynamics in storytelling? \textbf{RQ4}: How do the interactions between participant gender, avatar gender, and dominant narrative actor gender impact the overall learning experience and cognitive engagement in a VR film production environment?

To address these questions, we design a VR-based film production task and conduct a 2$\times$2$\times$2 factorial experiment to analyze the impact of gendered avatars and narrative-driven content on 48 students' creative decisions and understanding of gender perspectives in a VR environment. By combining objective performance data (such as igroup presence questionnaire, standardized embodiment questionnaire, shot selection and lighting configurations) with subjective feedback collected through surveys and interviews, {we aim to explore the impact of such a virtual environment on student engagement,} learning outcomes, and understanding of gender perspectives. The findings will provide valuable insights for educators, helping to design more inclusive and effective VR educational tools, particularly for enhancing film production training and deepening the understanding of gender dynamics in film narratives.

\section{Related Work}

\paragraph{{Immersive VR for Film Training}}
{An earlier work on VR film training reported clear efficiency gains: students who could put themselves into a virtual set adjusted lights, cameras, and blocking far more often than on a physical stage, accelerating skill acquisition while saving studio space and equipment wear \cite{wei2023feeling}. Presence theory explains this effect. Spatial presence supports attention and motor learning \cite{slater1997framework}, whereas self-presence boosts agency and confidence \cite{lee2004presence}; considering these factors in a bundle could result in predicting higher motivation and better understanding of real-world practice.} {Second-generation platforms now replicate the full production pipeline: simulated sound recording, multi-user communication, virtual cameras, and HMD-based waveform/vector scopes let learners refine coverage in minutes rather than hours \cite{xu2024transforming,wei2024hearing,wei2025illuminating,xu2023cinematography}. Two gaps remain: (1) Gender-matched avatars by default -- Participants are treated as a homogeneous pool, always given an avatar that matches their own gender. (2) Missing creative metrics -- Prior work logs task completion or speed, not stylistic choices such as gaze or lighting quality.} {We therefore investigate whether gendered presence alters style. When male and female learners inhabit male- or female-coded avatars in the same virtual studio, do they prefer soft keys over hard backlights, emotional close-ups over kinetic wides? Our work, from the angle of VR film pedagogy, suggests alternative views of abandoning one-size-fits-all avatars and embracing varied embodiment to broaden perspectives and reduce gender bias.}

\paragraph{{Avatar Congruence and Self-Presence}}
{Extensive embodiment work shows that when an avatar mirrors a user's salient traits—body size, skin tone, gender—self-presence (``this virtual body is me'') rises, sensorimotor mapping shortens, and task fluency improves \cite{do2024cultural,freeman2021body,kim2024self,kim2023or,buck2022impact,wei2025towards}. Such congruence has boosted learning in assembly, surgery, and architectural walk-throughs, suggesting that a matching gendered avatar should likewise streamline cinematography practice.} {Yet incongruence can act as a creative catalyst. Peck \& Gonzalez-Franco \cite{peck2021avatar} found that race-swap avatars heightened agency and critical self-reflection because users watched their virtual bodies more closely. Do et al. \cite{do2024cultural} reported that students with non-matching avatars checked mirrors significantly more often, signalling deeper meta-cognition. In a storytelling context, this heightened self-monitoring may push learners to question habitual framing rules—cross-gender embodiment could, for example, nudge men toward ``female-gaze'' close-ups and women toward ``male-gaze'' kinetic wides.}


\paragraph{{Proteus Effect and Gendered Avatars}}
{The \textit{Proteus Effect} shows that users' attitudes and behaviours drift toward salient avatar cues \cite{yee2007proteus,szolin2023exploring,banakou2018virtually,guegan2016avatar}. Follow-up work has exposed strong gender-swap consequences: men using female avatars reduced verbal dominance, whereas women using male avatars became more assertive, an effect magnified by pre-existing gender stereotypes \cite{sherrick2014role}; women inhabiting hyper-sexualised female avatars reported higher self-objectification and body anxiety \cite{fox2013embodiment}; cartoon-styled female avatars triggered stronger objectification judgments than equivalent male avatars \cite{nowak2015inferences}. A meta-analysis confirms small-to-medium behavioural shifts across multiple avatar traits, including gender cues \cite{ratan2020avatar}. Yet nearly all studies examined brief social or game contexts.}

\paragraph{{Gendered Gaze and Narrative Power}}

{Feminist film theory argues that mainstream cinema encodes a male gaze that objectifies women and centres heterosexual male desire \cite{smelik2007feminist,evans1995gaze}; later scholars map a female gaze grounded in empathy, reciprocity and character subjectivity \cite{boler1997risks,hollinger2012feminist,dirse2013gender} and note directorial gender asymmetries in depicting female desire \cite{doane2013woman}. In VR, first-person immersion can intensify these hierarchies, as gaze tracking shows that male users linger longer on sexualised female bodies \cite{fox2009virtual}, or subvert them by enabling cross-gender or queer embodiment that rehearses alternative identities \cite{freeman2022acting}. Design studies reveal that avatar form interacts with sexism: cartoonish female avatars prompt stronger objectification than male equivalents \cite{nowak2015inferences}; hardware/content defaults still privilege male anthropometrics, producing comfort gaps for women \cite{stanney2020virtual}. Han et al. extend this discussion with a large-scale longitudinal study in the metaverse, demonstrating that avatar design and environmental context jointly steer group norms over weeks, not minutes, evidence that narrative setting and embodiment interact over time \cite{han2023people}. To disentangle visual dominance from avatar cues, we recreated the iconic \textit{Black Swan} pas de deux, often cited as a textbook male-gaze sequence, in two matched versions where either the male or the female dancer initiates the action, controls the lifts, and holds the center of the frame \cite{gibson2011disciplining}. By logging light ratios, shot scales, and gaze paths, our 2$\times$2$\times$2 design provides the first test of how avatar embodiment and narrative power jointly sculpt creative decisions in VR film production.}

\section{Methods}
To investigate how gendered avatars and gaze dynamics in virtual environments influence students' understanding of gender perspectives in film production, our experiment incorporated two learner genders (male and female), two types of avatars used by the learners (male and female avatars), and two dominant narrative actor gender combinations (male and female actor-driven narrative), resulting in a 2$\times$2$\times$2 between-subjects experimental design.

\begin{figure}[ht]
    \centering
    \includegraphics[width=8cm]{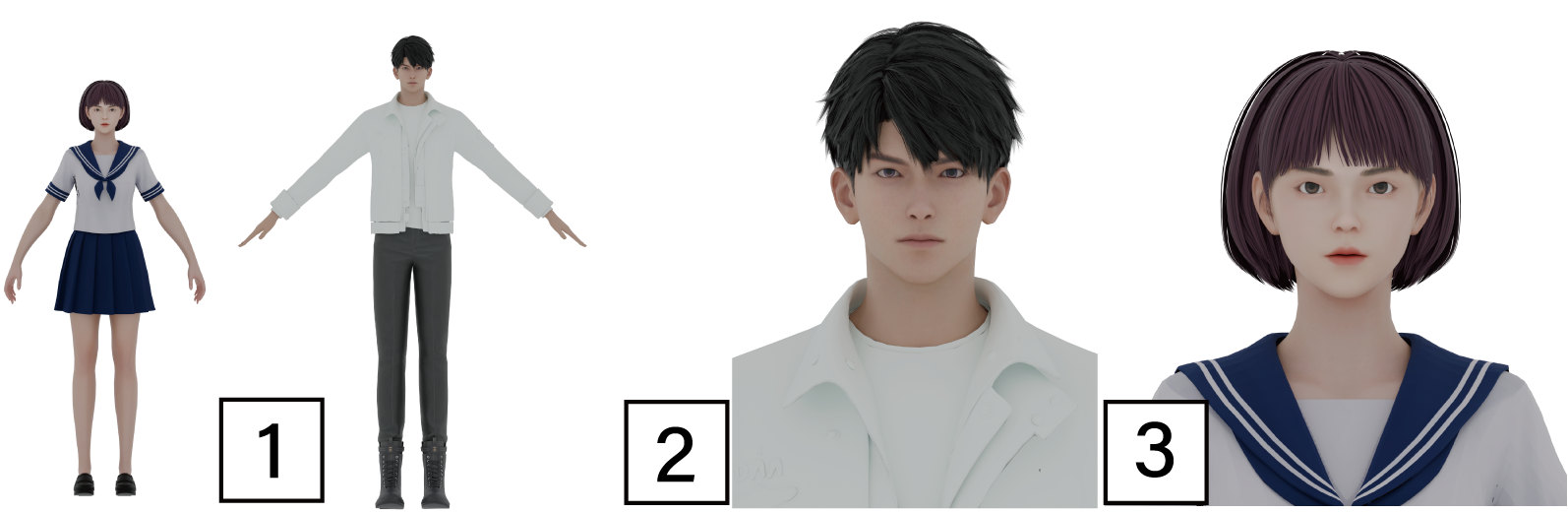}
    \caption{(1) Two avatar types; (2) Male avatar; (3) Female avatar.}
    \label{avatar}
\end{figure}

Participants engaged in activities with two different avatar types (Figure \ref{avatar} - (1)): male and female avatars (Figures \ref{avatar} - (2) and \ref{avatar} - (3)). These two avatar types were selected to explore how different gendered avatars influence users' sense of self-presence and, in turn, impact their self-perception. Given that all our participants are of East Asian descent, we designed the avatars with typical East Asian features, including skin tone, height, hair color, and facial features. Both the male and female avatars were carefully designed with the same visual quality to ensure consistency across conditions.

\begin{figure}[ht]
    \centering
    \includegraphics[width=8cm]{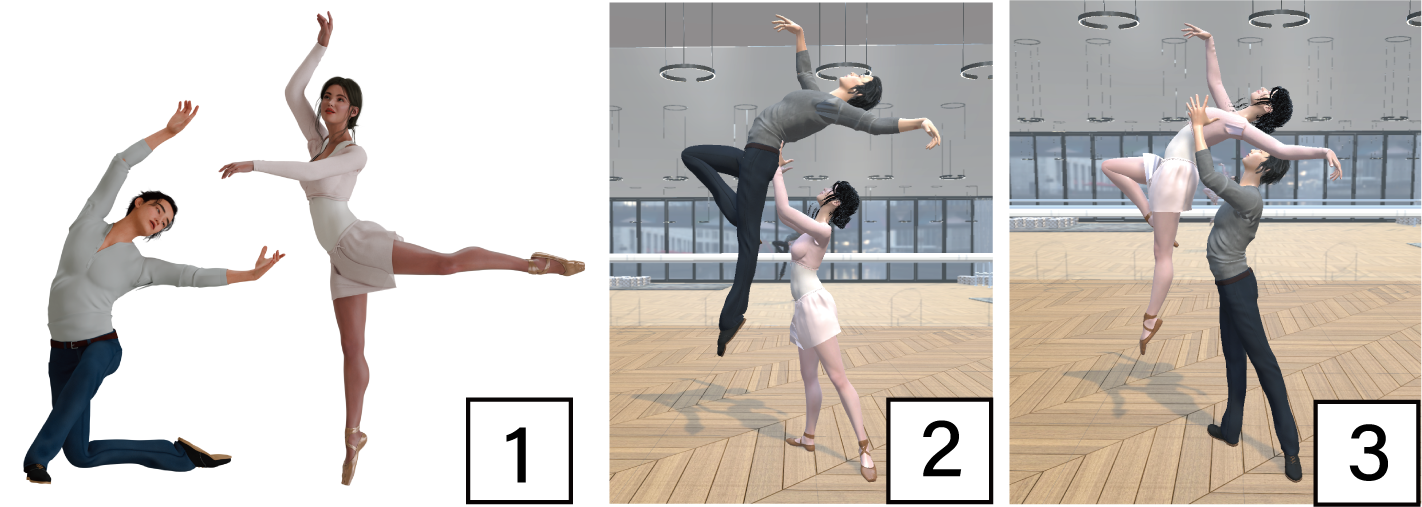}
    \caption{(1) Two actor types; (2) Female actor-driven narrative; (3) Male actor-driven narrative.}
    \label{actor}
\end{figure}

The gender of the dominant narrative actor is another independent variable, with two conditions: a male actor-driven narrative and a female actor-driven narrative, as shown in Figure \ref{actor} (1) - (3). The primary distinction between these conditions lies in the allocation of agency between the actor characters, while other elements such as dialogue content, performance actions, and costumes remain consistent. We chose to film a dance scene inspired by the iconic dance shots from the film \textit{Black Swan}, which is considered a representative example of the male gaze \cite{gibson2011disciplining}. Critics often point to the film's visual and narrative strategies, which emphasize a voyeuristic perspective on the female protagonist's body and emotional journey. The camera frequently lingers on her physical form and psychological fragility, framing her through a lens that accentuates her sexualization and bodily transformation rather than her subjective interiority. This aligns closely with classic theories of the male gaze, which posit that female characters are habitually presented as objects to be observed rather than as fully realized agents. By referencing \textit{Black Swan}, we underscore how male-dominant and female-dominant narrative perspectives can yield distinctly different power dynamics and modes of spectatorship. Therefore, we set up two narrative conditions to showcase the male and female gaze perspectives through actions alone. In the male actor-driven narrative scenario, the male actor initiates the actions, while the female actor passively responds and follows his movements (i.e., male-dominant, female-passive). Conversely, in the female actor-driven narrative scenario, the female actor leads, and the male actor passively follows her actions (i.e., female-dominant, male-passive). In both conditions, participants completed a film production task and created a 45-second short video.  

To ensure the accuracy and consistency of the experimental results, we provided each participant with the same experimental materials and filming objectives, minimizing the potential impact of task difficulty on the outcomes. In each case, all participants used the same equipment and virtual reality setup to eliminate any influence on the results caused by hardware limitations or technical issues. Furthermore, aside from differences in participant gender, avatar gender, and the gender of the dominant narrative actor, the virtual scenes used in the tasks remained consistent in terms of filming equipment (lighting, cameras), environment (shooting locations and actor costumes), actor performances, and the tools and resources available to the participants.

\subsection{Research Hypotheses} \label{Research Hypotheses}

{Prior research converges on four interrelated strands that underpin our hypotheses. (i) Immersive VR film-training studies demonstrate that elevated spatial and self-presence substantially accelerate the acquisition of lighting and camera skills \cite{wei2024hearing,wei2025illuminating,xu2023cinematography}. (ii) Avatar–user gender congruence further shortens sensorimotor mapping and enhances task fluency, whereas gender-incongruent embodiment heightens self-monitoring and often provokes exploratory, non-canonical stylistic choices \cite{do2024stepping,freeman2021body}. (iii) These embodiment effects accord with the Proteus paradigm: behaviour shifts toward salient avatar cues, and gender-swap experiments consistently document measurable reversals in dominance and assertiveness \cite{sherrick2014role,ratan2020avatar}. (iv) Feminist film theory distinguishes an objectifying male gaze from an empathic female gaze \cite{smelik2007feminist,dirse2013gender}; complementary VR studies document systematic gender biases in visual attention, while longitudinal metaverse research shows that sustained alignment of avatar and narrative cues consolidates shared creative norms \cite{fox2009virtual,han2023people}. Collectively, these strands imply that the specific alignment or misalignment of participant gender, avatar gender, and narrative dominance will systematically shape presence, shot selection, and gaze dynamics in VR film production, thereby motivating our four hypotheses.} \textbf{H1:} Participants using gender-congruent avatars (i.e., male participants using male avatars, female participants using female avatars) will report higher levels of presence and perception compared to participants using gender-incongruent avatars. \textbf{H2:} Female participants using a female avatar are more likely to use soft lighting and background rendering to emphasize emotional interactions between characters, while male participants using a male avatar are more likely to use action shots and strong visual contrasts to highlight the dynamic and tension of the scene. \textbf{H3:} Male participants using a female avatar are more likely to choose shots related to emotional interaction, reducing focus on action and body details, while female participants using a male avatar are more likely to choose shots that emphasize visual impact and action scenes. \textbf{H4:} There will be an interaction effect between participant gender, avatar gender, and actor-driven narrative gender, such that the most positive outcomes in terms of understanding gender perspectives will occur when participants' gender aligns with both their avatar gender and the dominant actor gender.

\subsection{Film DanceStudio}
\begin{figure}[ht]
    \centering
    \includegraphics[width=8cm]{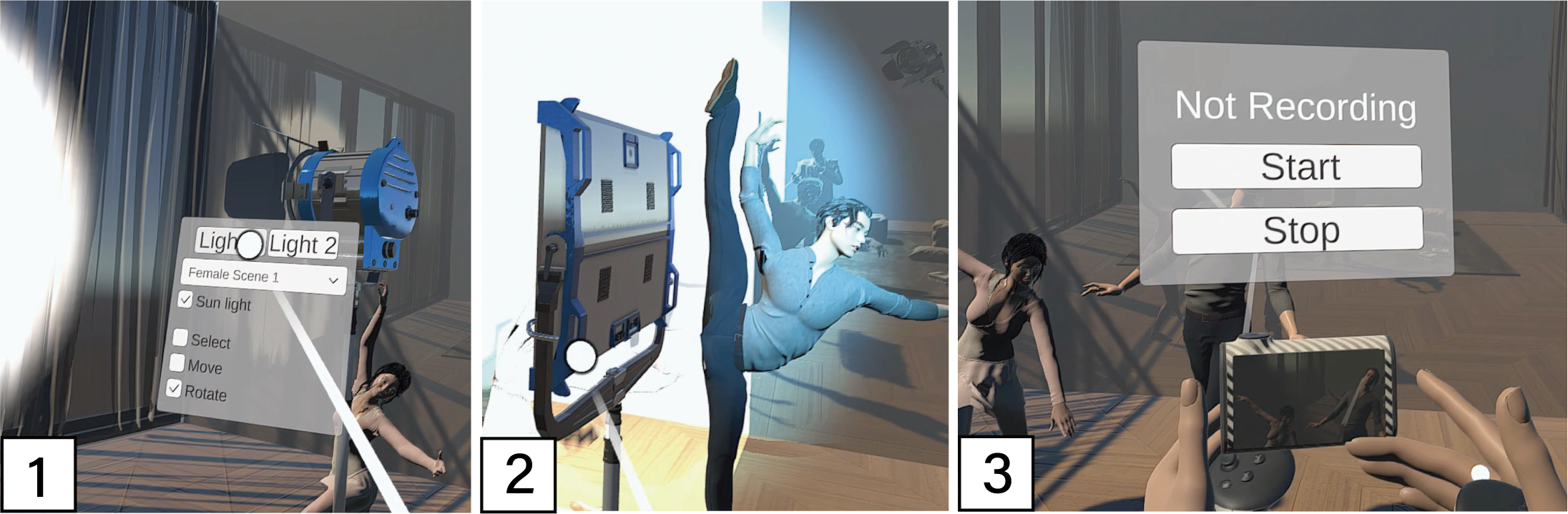}
    \caption{(1) Tungsten Light; (2) LED Light; (3) Handheld camera supporting VR recording.}
    \label{camear}
\end{figure}

To validate our hypotheses, we developed \textit{Film DanceStudio}, a VR system grounded in presence theory and the \textit{Proteus Effect}. \textit{Film DanceStudio} simulates key aspects of film production—such as cinematography and lighting—within an interactive virtual studio. In this environment, participants can experiment with various lighting devices (tungsten, LED; see Figure \ref{camear}-(1) - (2)), use a handheld VR camera with in-headset recording (see Figure \ref{camear}-(3)), and select from two avatar genders (male, female; see Figure \ref{avatar}) and two narrative role conditions (male-led, female-led; see Figure \ref{actor}). Real-world filmmaking practices closely mirror these configurations, combining multiple lighting setups and tailoring creative decisions to specific production goals. 

\begin{figure}[ht]
    \centering
    \includegraphics[width=6cm]{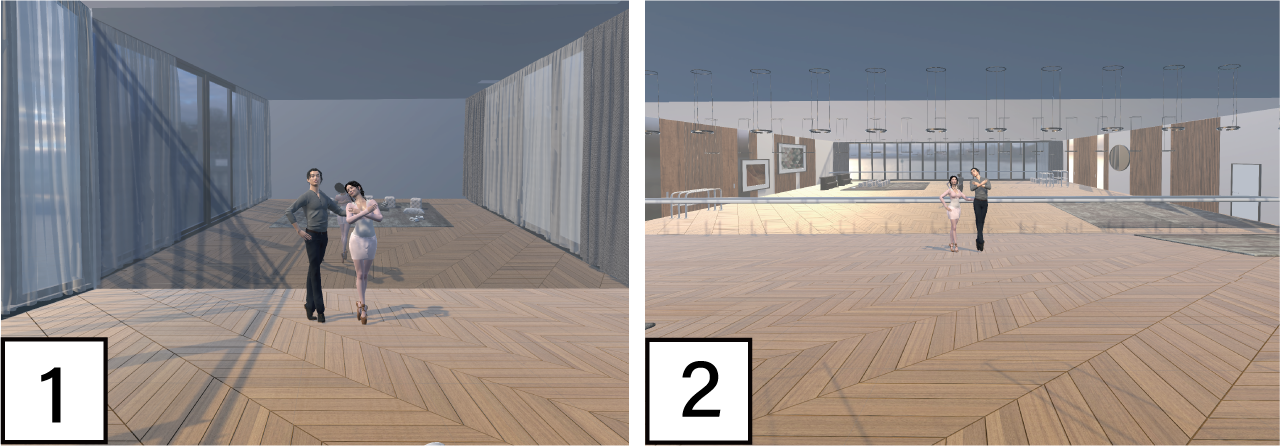}
    \caption{Both scenes are equipped with large floor-to-ceiling mirrors, allowing users to observe themselves at all times: (1) Training Scene; (2) Shooting Scene.}
    \label{scence}
\end{figure}

Users first learn lighting and camera operation in a training scene, where they also adapt to observing and interacting with their avatar and actor; see Figure \ref{scence}-(1). This process is expected to take 15 minutes. Subsequently, users transition into the shooting scene for formal creation, with this phase expected to take 30 minutes; see Figure \ref{scence}-(2). Both scenes are equipped with large floor-to-ceiling mirrors, allowing users to observe themselves at all times.

By analyzing the creative behaviors of participants under different conditions, we aim to explore gaze dynamics, specifically the interaction between gender, avatars, and camera perspective, and deepen our understanding of how male and female viewpoints change within VR environments. This flexibility, alongside various equipment configurations, enhances user engagement and immersion, thereby offering valuable insights for the design of future VR-based teaching systems.

\subsubsection{Film DanceStudio Interaction Pipeline}

\begin{figure}[ht]
    \centering
    \includegraphics[width=8cm]{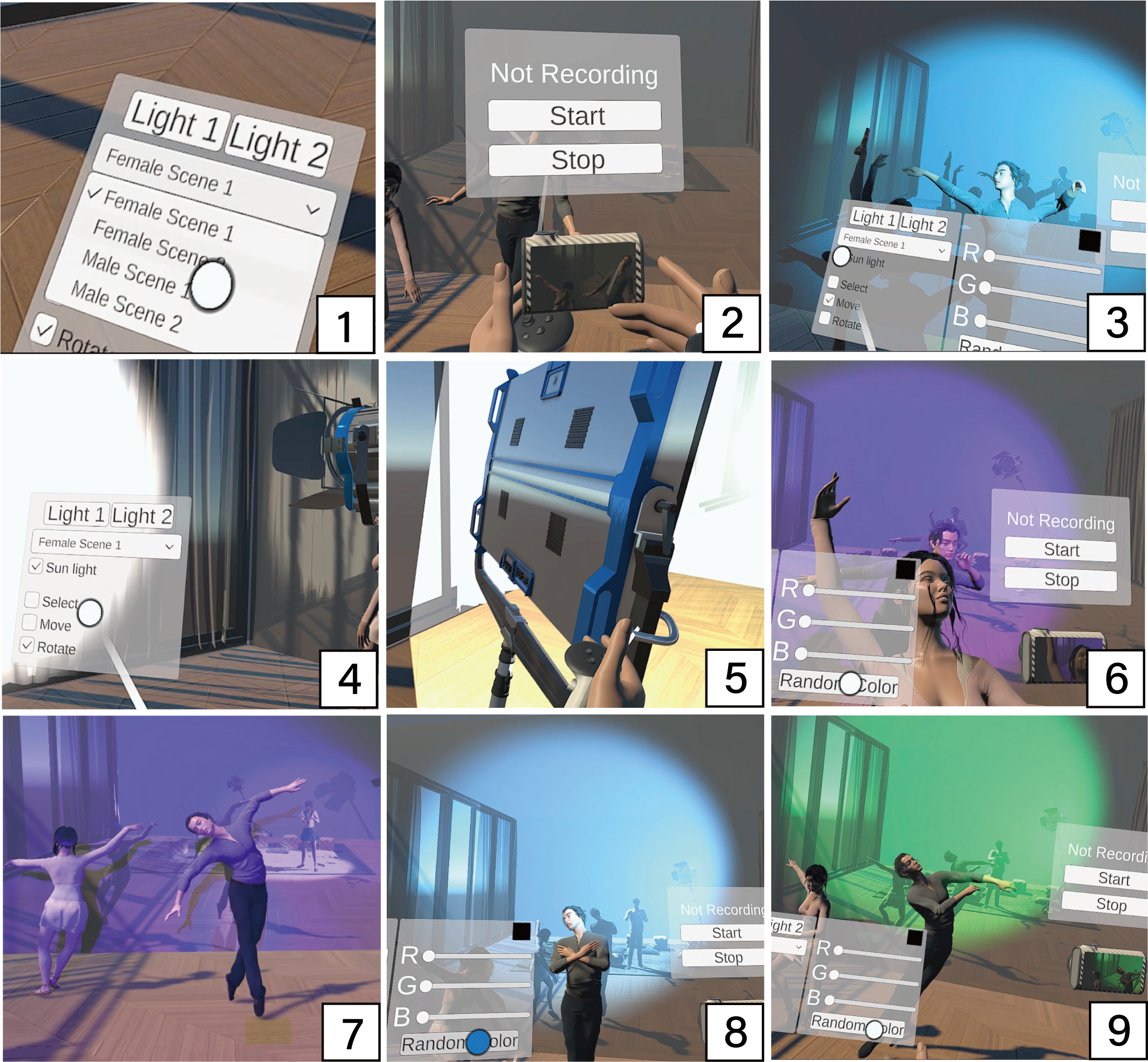}
    \caption{(1) Avatar selection and Role condition setup. (2) Virtual camera recording control. (3) Ambient light toggle for lighting effects. (4) Select, move, and rotate buttons for object and light adjustments. (5) Two types of lighting fixtures available. (6) Lighting control panel with RGB color settings. (7) Random light color generation through RGB control. (8) Simultaneous operation of multiple lights for complex designs. (9) Select, grab, and move lights, cameras, and props.}
    \label{pipeline}
\end{figure}

The interactive pipeline provides a user-friendly interface that helps students focus on learning filmmaking. After confirming their avatar and entering the login interface, users can choose their actor narrative role conditions, as shown in Figure \ref{pipeline} - (1). Users can operate the virtual camera and press the record button to start video recording, as demonstrated in Figure \ref{pipeline} - (2). Additionally, users can toggle the sunlight on or off to observe the effects of their lighting setup, as shown in Figure \ref{pipeline} - (3). Above the left controller, a virtual panel contains buttons for selecting, moving, and rotating objects. These buttons correspond to object selection, movement, and the adjustment of light tilt angles, as shown in Figure \ref{pipeline} - (4). In the \textit{Film DanceStudio} system, users have access to two types of lighting, as illustrated in Figure \ref{pipeline} - (5). When users approach any lighting fixture, the dashboard automatically appears to facilitate better interaction with lighting controls, offering a range of adjustable parameters. For both lighting types, a standardized control panel is available to adjust RGB color settings, allowing users to freely generate the desired light colors, as shown in Figure \ref{pipeline} - (6). Students can observe their avatar and actions in real-time through a mirror positioned in front of them, as depicted in Figure \ref{pipeline} - (7). The system supports the simultaneous operation of multiple lights, enabling students to create complex lighting designs, as shown in Figures \ref{pipeline} - (8) and (9).

\subsection{Dependent Variables}
We primarily measured participants' self-identification and sense of immersion within the virtual film production environment, as well as the effectiveness of the resulting film. Accordingly, two questionnaires were distributed after the completion of each experimental condition, as described below. Additionally, participants took part in a 10-minute semi-structured interview to gather qualitative feedback.

\paragraph{Igroup Presence Questionnaire}
To assess the participants' sense of presence during the experiment, we used the Igroup Presence Questionnaire \cite{schubert2001experience}. We chose this questionnaire because it has been extensively validated and widely applied in measuring presence across various domains. Among the many dimensions, we extracted only four main dimensions, each rated on a 7-point Likert scale: ``Spatial Presence,'' ``Involvement,'' ``Realism,'' and ``Control'' from the IPQ. Together, these dimensions provide a comprehensive perspective on immersive virtual experiences. 

\paragraph{Standardized Embodiment Questionnaire}
We used the Standardized Embodiment Questionnaire (SEQ) \cite{peck2021avatar} to assess users' embodiment experiences in virtual environments. This scale has been widely used in previous research to quantify the extent to which users feel that the virtual body is part of their own body and to assess their interaction with the virtual body. The questionnaire uses a 7-point Likert scale. Among the many dimensions, we extracted only four main dimensions: ``Body Sensation,'' ``Sense of Control,'' ``Awareness,'' and ``Identity with the Virtual Body'' from the SEQ.  These items aim to comprehensively evaluate users' perception, control, identity, and presence in the virtual body, providing a standardized tool for understanding embodiment in virtual experiences.

\paragraph{Observation and Semi-structured interviews}
During the experiment, participants performed the task inside a VR headset while we mirrored the headset’s view to a desktop monitor, allowing us to observe their behaviour in real time. At the end of the experiment, participants engaged in semi-structured interviews to provide detailed feedback on their filming tasks. Participants were asked three key questions focusing on their experiences and outcomes of the filming task, including: 1) How did the type of embodiment influence your ability to perform the filming task? 2) What are your feelings and views on the gender of the dominant narrative actor avatar in the virtual environment? 3) How did this affect your overall filming experience?

\begin{figure}[ht]
    \centering
    \includegraphics[width=8.5cm]{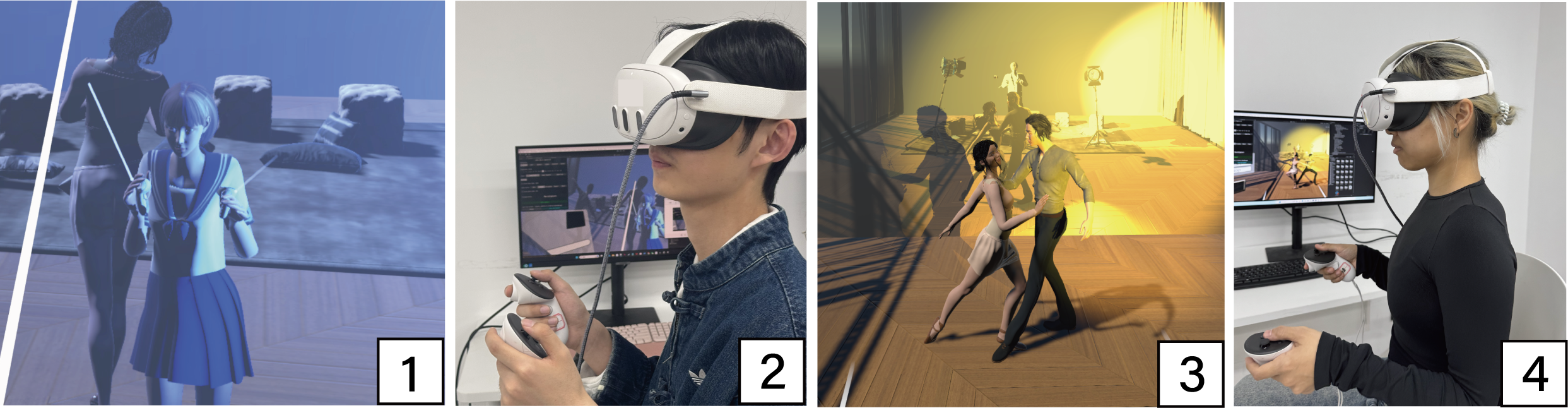}
    \caption{(1)-(2) A male user is using a female avatar to shoot a male-led scene (Group C). (3)-(4) A female user is using a male avatar to shoot a male-led scene (Group E).}
    \label{Apparatus}
\end{figure}

\paragraph{Apparatus}
The experiment was conducted in a quiet, spacious, and well-lit meeting room, as shown in Figure \ref{Apparatus}. Participants wore Meta Quest 3 headsets with 110\textdegree{} horizontal and 96\textdegree{} vertical field of view\footnote{\url{https://www.meta.com/quest/quest-3/}}. The VR environment runs on a computer and is streamed to the headsets in real-time using Unity 2022.3.33. The computer system used for this setup featured a 13th Gen Intel Core i9-13900K processor, 64 GB RAM, an NVIDIA RTX 4090 graphics card, and Windows 11. The source code is available at \url{https://anonymous.4open.science/r/589D}.

\subsection{Procedure}

\begin{figure}[ht]
    \centering
    \includegraphics[width=\linewidth]{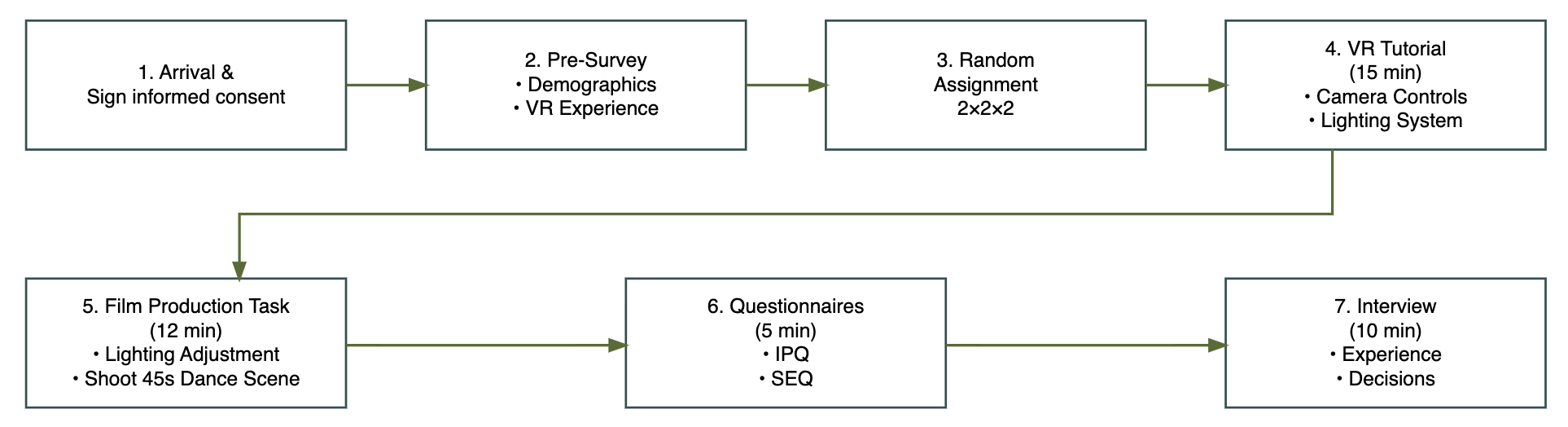}
    \caption{Flowchart of the experimental procedure.}
    \label{proc}
\end{figure}

The research protocol was approved by the Institutional Review Board (HSEARS20250325006) prior to the study. The in-person experimental session lasted approximately 45 minutes with a \$5 compensation. {The procedure comprised seven steps: (i) arrival and informed-consent signing; (ii) completion of a demographics and VR-experience survey (gender, prior VR use); (iii) random assignment to one of eight cells in the 2$\times$2$\times$2 design; (iv) a 15-min VR tutorial inside the virtual studio covering camera controls, lighting rigs, and task goals; (v) a 12-min Film Production Testing Task in which the participant, embodied in the assigned avatar, shot a 45-s dance scene with self-selected framing, movement, and lighting (actions mirrored to an external display); 
(vi) completion of IPQ and SEQ questionnaires (5 min); (vii) a 10-min semi-structured interview probing avatar experience and shot decisions. Figure~\ref{proc} illustrates the experimental procedure.

All participants followed the same 15-min VR tutorial on camera controls, lighting rigs, and task goals before filming, and they used identical tools throughout. Therefore, any observed performance variation can be attributed to the participant, avatar, and narrative genders rather than to differences in task complexity.

\paragraph{Film Production Testing Task}
This task aims to understand how gender perspectives influence film production through immersive virtual experiences. In this task, participants are required to create a 45-second short film within a virtual environment, focusing on how they choose camera angles and shots that reflect gender dynamics and gaze characteristics. Participants interact with two types of avatars (male and female) and two narrative actor conditions (male-driven and female-driven) to explore the impact of gendered avatars and actor narratives on film production choices. 

The final evaluation is conducted by two experts, who assess the films based on shot types (e.g., wide-angle vs. close-up), lighting configurations (e.g., high-contrast vs. soft lighting), special shots (e.g., action vs. emotional), and overall style. They independently coded each clip's shot scale, lighting style, and overall gaze category (male, female, neutral) and reconciled their ratings. Each shot is classified according to the male gaze, female gaze, or neutral gaze. Male gaze is typically associated with emphasizing body details and visual impact \cite{oliver2017male,cooper2000chick}, while the female gaze usually focuses on emotional investment and subtle, intimate interactions \cite{dirse2013gender,langga2020female}. The neutral gaze strives for balance and avoids stereotypical gender portrayals. The test considers the following factors: 1) Shot and lighting types. Male gaze characteristics are reflected in action-driven shots, such as fast-moving or wide-angle shots, combined with high-contrast lighting, which emphasizes the visual impact of the scene and story. In contrast, the female gaze is characterized by close-up shots and soft lighting that highlight emotional exchanges, intimacy, or subtle emotional details between characters, focusing on their inner emotional states and subtle interactions rather than purely visual impact. Female gaze shots emphasize characters' emotional states and psychological changes, avoiding excessive gendered or objectified portrayals, and focusing more on the emotional connection between people. 2) Participants' perspective on the performers. In male gaze perspectives, participants tend to focus more on body parts (e.g., legs, chest, hips), emphasizing visual impact. From a female gaze perspective, participants are more likely to capture the performers' facial expressions and hand movements, emphasizing emotional interaction.

The experts come from different backgrounds, which uniquely qualify them for this task. Expert 1 (female), a senior director with 10 years of experience, has a deep understanding of perspective and gaze, particularly in narrative-driven film production. Expert 2 (male), a film cinematographer with 10 years of experience, brings expertise in shot composition, lighting techniques, and how these elements contribute to visual storytelling. Their diverse perspectives help assess both the technical aspects and the gendered dynamics present in the video, providing a comprehensive evaluation.

\subsection{Participants}
We recruited 48 participants via social media, who provided self-reported demographic information, including gender (24 females, 24 males) and age ($\bar{M}$ = 24.3, SD = 2.95). All participants were cisgender, with the absence of non-binary gender participants. 41 participants reported prior VR experience, and all participants had normal or corrected vision. Participants were assigned to one of eight experimental groups based on user gender, avatar gender, and actor-driven narrative, as shown in Table \ref{group}. 
A computer script performed block randomisation stratified by participant gender: each new enrollee was placed in the least-filled cell for their gender, in blocks of eight, yielding exactly six participants per condition.

\begin{table}[ht]
\center
\caption{Participants were assigned to one of eight groups (A-I) based on user gender (M/F), avatar gender (M/F), and actor-driven narrative (M/F). The groups are: Group A (M/M/M), Group B (M/M/F), Group C (M/F/M), Group D (M/F/F), Group E (F/M/M), Group F (F/M/F), Group G (F/F/M), and Group H (F/F/F), each representing a unique combination of these factors.}
\scalebox{0.7}{
\begin{tabular}{ccccccccc}
\hline
                                                                & A & B & C & D & E & F & G & H \\ \hline
\begin{tabular}[c]{@{}c@{}}User\\ Gender\end{tabular}           & M & M & M & M & F & F & F & F \\
\begin{tabular}[c]{@{}c@{}}Avatar \\ Type\end{tabular}          & M & M & F & F & M & M & F & F \\
\begin{tabular}[c]{@{}c@{}}Actor-Driven \\ Narrative\end{tabular} & M & F & M & F & M & F & M & F \\ \hline
\end{tabular}
\label{group}
}
\end{table}

\subsection{Data Analysis Approach}
Following the guidelines provided in the literature \cite{do2024stepping,do2024cultural}, we first calculated the average scores and the overall total scores for all subscale questions. Next, we conducted the Shapiro-Wilk test to assess the normality of the score distributions \cite{razali2011power}. Since the results indicated that several measurement variables exhibited non-normal distributions, we opted to use the Aligned Rank Transform (ART) method for a 2$\times$2$\times$2 factorial ANOVA \cite{wobbrock2011aligned}. With this approach, we examined the main effects and interaction effects of learner gender (male, female), avatar type (male avatar, female avatar), and the two dominant narrative actor gender patterns (male actor-driven narrative and female actor-driven narrative).

\section{Results}
We first discuss the results of the IPQ, followed by the SEQ, then report the expert ratings of the film production task, and finally, the results of the semi-structured interviews. We skip the reporting of these metrics IPQ (``Spatial Presence,'' ``Self-Presence,'' and ``Control.'') and SEQ (`Sense of Control,'' and ``Identity with the Virtual Body.''), due to the absence of statistical significance. Table \ref{table1} lists the mean values and standard deviations of all metrics being measured.

\begin{table}[ht]
\centering
\caption{This table summarizes the mean (M) and standard deviation (SD) of the questionnaire results (1 (worst) to 7 (best)). Significant interaction effects are denoted as follows: an asterisk (*) for User Gender and Avatar Gender, a plus sign (+) for User Gender and Actor-Driven Narrative, an equals sign (=) for Avatar Gender and Actor-Driven Narrative, and a hash symbol (\#) for a three-way interaction among User Gender, Avatar Gender, and Actor-Driven Narrative.}
\scalebox{0.75}{
\begin{tabular}{ccclc}
\hline
                                                   & \begin{tabular}[c]{@{}c@{}}Group A\\ (MMM)\\ M(SD)\end{tabular} & \begin{tabular}[c]{@{}c@{}}Group B\\ (MMF)\\ M(SD)\end{tabular} & \multicolumn{1}{c}{\begin{tabular}[c]{@{}c@{}}Group C\\ (MFM)\\ M(SD)\end{tabular}} & \begin{tabular}[c]{@{}c@{}}Group D\\ (MFF)\\ M(SD)\end{tabular} \\ \hline
IPQ                                                &                                                                 &                                                                 & \multicolumn{1}{c}{}                                                                &                                                                 \\ \hline
Involvement*+                                      & 3.67(1.97)                                                      & 4.17(1.33)                                                      & \multicolumn{1}{c}{4.83(1.17)}                                                      & 5.33(1.86)                                                      \\
Spatial Presence                                   & 5(1.55)                                                         & 4.67(1.21)                                                      & \multicolumn{1}{c}{5.5(0.84)}                                                       & \multicolumn{1}{l}{5.83(1.47)}                                  \\
Control                                            & \multicolumn{1}{l}{4.67(2.07)}                                  & \multicolumn{1}{l}{4.83(0.98)}                                  & 5.17(1.72)                                                                          & 5.5(2.35)                                                       \\
Self-Presence                                      & \multicolumn{1}{l}{4.67(1.97)}                                  & 5.5(0.84)                                                       & 5.33(0.52)                                                                          & \multicolumn{1}{l}{6.17(1.60)}                                  \\ \hline
SEQ                                                &                                                                 &                                                                 & \multicolumn{1}{c}{}                                                                &                                                                 \\ \hline
Awareness+\#                                       & 2.33(1.63)                                                      & 4.33(1.51)                                                      & \multicolumn{1}{c}{4.33(1.51)}                                                      & 3(1.1)                                                          \\
Body Sensation*=                                   & 2.5(1.64)                                                       & 4.83(1.17)                                                      & \multicolumn{1}{c}{4.33(1.21)}                                                      & 3.83(2.14)                                                      \\
Sense of Control                                   & \multicolumn{1}{l}{3.67(1.75)}                                  & 5(1.10)                                                         & 5.33(1.75)                                                                          & 4(2)                                                            \\
\multicolumn{1}{l}{Identity with the Virtual Body} & 2.5(1.64)                                                       & 4(1.26)                                                         & 4.67(1.03)                                                                          & 3(1.10)                                                         \\ \hline
                                                   & \begin{tabular}[c]{@{}c@{}}Group E\\ (FMM)\\ M(SD)\end{tabular} & \begin{tabular}[c]{@{}c@{}}Group F\\ (FMF)\\ M(SD)\end{tabular} & \multicolumn{1}{c}{\begin{tabular}[c]{@{}c@{}}Group G\\ (FFM)\\ M(SD)\end{tabular}} & \begin{tabular}[c]{@{}c@{}}Group H\\ (FFF)\\ M(SD)\end{tabular} \\ \hline
IPQ                                                & \multicolumn{1}{l}{}                                            & \multicolumn{1}{l}{}                                            &                                                                                     & \multicolumn{1}{l}{}                                            \\ \hline
Involvement*+                                      & 5.17(0.98)                                                      & 4.5(1.22)                                                       & \multicolumn{1}{c}{5(1.55)}                                                         & 3.67(1.21)                                                      \\
Spatial Presence                                   & \multicolumn{1}{l}{5.67(0.82)}                                  & 5.5(1.38)                                                       & 5.17(0.98)                                                                          & \multicolumn{1}{l}{5.17(0.75)}                                  \\
Control                                            & 4.5(2.43)                                                       & 5.5(1.05)                                                       & 5.33(1.03)                                                                          & 4.5(1.05)                                                       \\
Self-Presence                                      & \multicolumn{1}{l}{5.33(0.82)}                                  & \multicolumn{1}{l}{5.67(1.03)}                                  & 5.33(1.03)                                                                          & 5(1.10)                                                         \\ \hline
SEQ                                                &                                                                 &                                                                 &                                                                                     & \multicolumn{1}{l}{}                                            \\ \hline
Awareness+\#                                       & 3.83(1.94)                                                      & 2.17(1.94)                                                      & \multicolumn{1}{c}{4.5(1.38)}                                                       & 3.17(1.47)                                                      \\
Sense of Control                                   & 4.5(1.76)                                                       & 3.5(2.43)                                                       & 5.33(1.21)                                                                          & 4.5(1.64)                                                       \\
\multicolumn{1}{l}{Identity with the Virtual Body} & \multicolumn{1}{l}{3.83(1.47)}                                  & \multicolumn{1}{l}{3.33(2.88)}                                  & 3.83(1.17)                                                                          & \multicolumn{1}{l}{2.67(1.51)}                                  \\
Body Sensation*=                                   & 4.5(1.05)                                                       & 3.83(1.47)                                                      & \multicolumn{1}{c}{5.33(0.82)}                                                      & 3.33(1.63)                                                      \\ \hline
\end{tabular}
}
\label{table1}
\end{table}

\subsection{Igroup Presence Questionnaire}

The Igroup Presence Questionnaire revealed distinct patterns across the measured variables, as illustrated in Figure \ref{Q} (1)-(2). Our statistical analysis shows a significant interaction effect between user gender and actor-driven narrative on users' sense of presence ($F_{1,40}$ = 5.83, p = 0.0203, $\text{partial }\eta^{2}=.13$). This indicates that the effect of narrative conditions on presence varies by user gender, meaning that male and female users may respond differently to actor-driven narratives. We found that males in female-dominated narratives (or females in male-dominated narratives) reported higher levels of presence. {A significant user--gender $\times$ avatar--gender interaction emerged ($F(1,40) = 4.13$, $p = 0.049$, $\text{partial }\eta^{2} = 0.09$), with both male and female learners reporting higher self-presence when their avatar's gender did
\emph{not} match their own. 

\begin{figure}[ht]
    \centering
    \includegraphics[width=8.5cm]{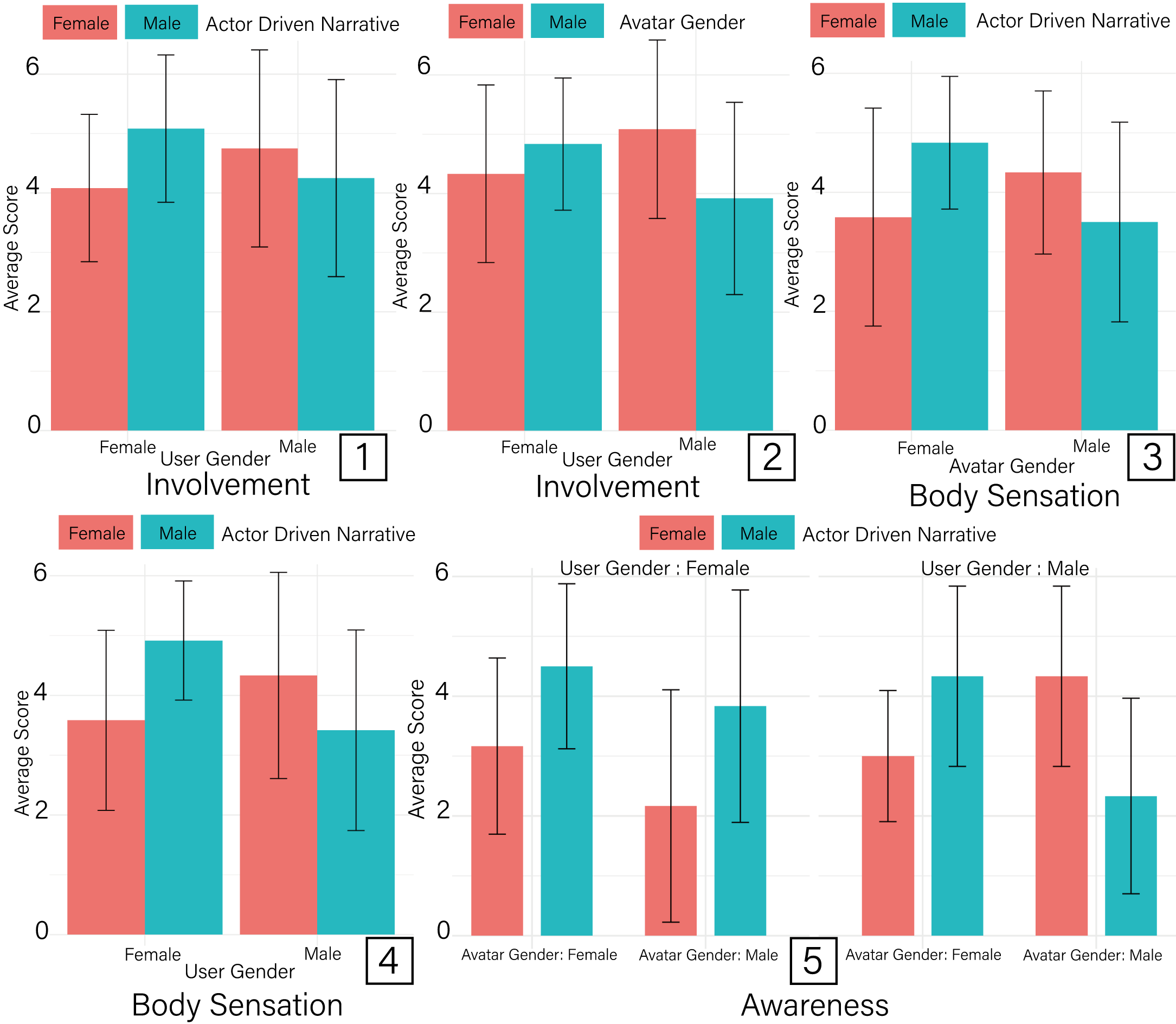}
    \caption{We provided all statistically significant results, along with the average scores from the IPQ and SEQ measurements. The charts compare the conditions across different user genders, avatar genders, and actor-driven narrative genders. Subfigures (1)-(2) display the IPQ scores for involvement (1 being the worst and 7 being the best). Subfigures (3)-(4) show the SEQ ratings for body sensation (1 being the worst and 7 being the best). Subfigure (5) illustrates the interaction effects of user gender, avatar gender, and actor-driven narrative gender on awareness in the SEQ.}
    \label{Q}
\end{figure}

\subsection{Standardized Embodiment Questionnaire}
The Standardized Embodiment Questionnaire also shows distinct patterns across the measured variables, as shown in Figure \ref{Q} (3)-(5), which shows only the items with significant differences. We observed two significant interaction effects on body sensation, as shown in Figure \ref{Q} (3)-(4). First, user gender moderated the impact of narrative condition, with a significant interaction between user gender and actor-driven narrative ($F_{1,40}$ = 6.77, p = 0.0129, {$\text{partial }\eta^{2}=0.15$}). 
Second, avatar gender also moderated the effect of narrative condition, as indicated by a significant interaction between avatar gender and actor-driven narrative ($F_{1,40}$ = 6.79, p = 0.0128, {$\text{partial }\eta^{2}= 0.15$}). 
This finding suggests that the influence of narrative structure on users' sense of body sensation differs depending on both the user's gender and the gender of the avatar. Specifically, male users in female-driven narratives (or female users in male-driven narratives) showed higher levels of body sensation. Additionally, users reported a stronger sense of body sensation when interacting with non-cisgender avatars. Statistical analysis of the data from the SEQ also revealed a significant interaction effect between the gender of the user, the gender of the avatar and the gender of the dominant narrative actor on awareness ($F_{1,40}$ = 4.28, p = 0.0451, {$\text{partial }\eta^{2} = 0.10$}), as shown in Figure \ref{Q} (5). 
Although the sample size was limited, the findings suggest that the interaction between user gender, avatar gender, and dominant narrative actor gender may provide valuable insights into the design of embodiment, gender identity, and creative decision-making in virtual reality. These findings potentially offer a new perspective for developing VR educational tools, especially when designing more immersive and gender-sensitive learning environments.

\subsection{Film Production Testing Task Results}\label{Film Production Testing Task Results}
In the Film Production Testing Task, we examined the performance of eight participant groups across different user genders, avatar genders, and dominant narrative actor genders. Figure \ref{gaze} (1)-(6) illustrates the visual effects of the ``Film Production Testing'' task conducted within the \textit{Film DanceStudio} environment, showcasing both the ``female gaze'' (Figure \ref{gaze} (2)(3)(5)) and the ``male gaze'' (Figure \ref{gaze} (1)(4)).

\begin{figure}[ht]
    \centering
    \includegraphics[width=8.5cm]{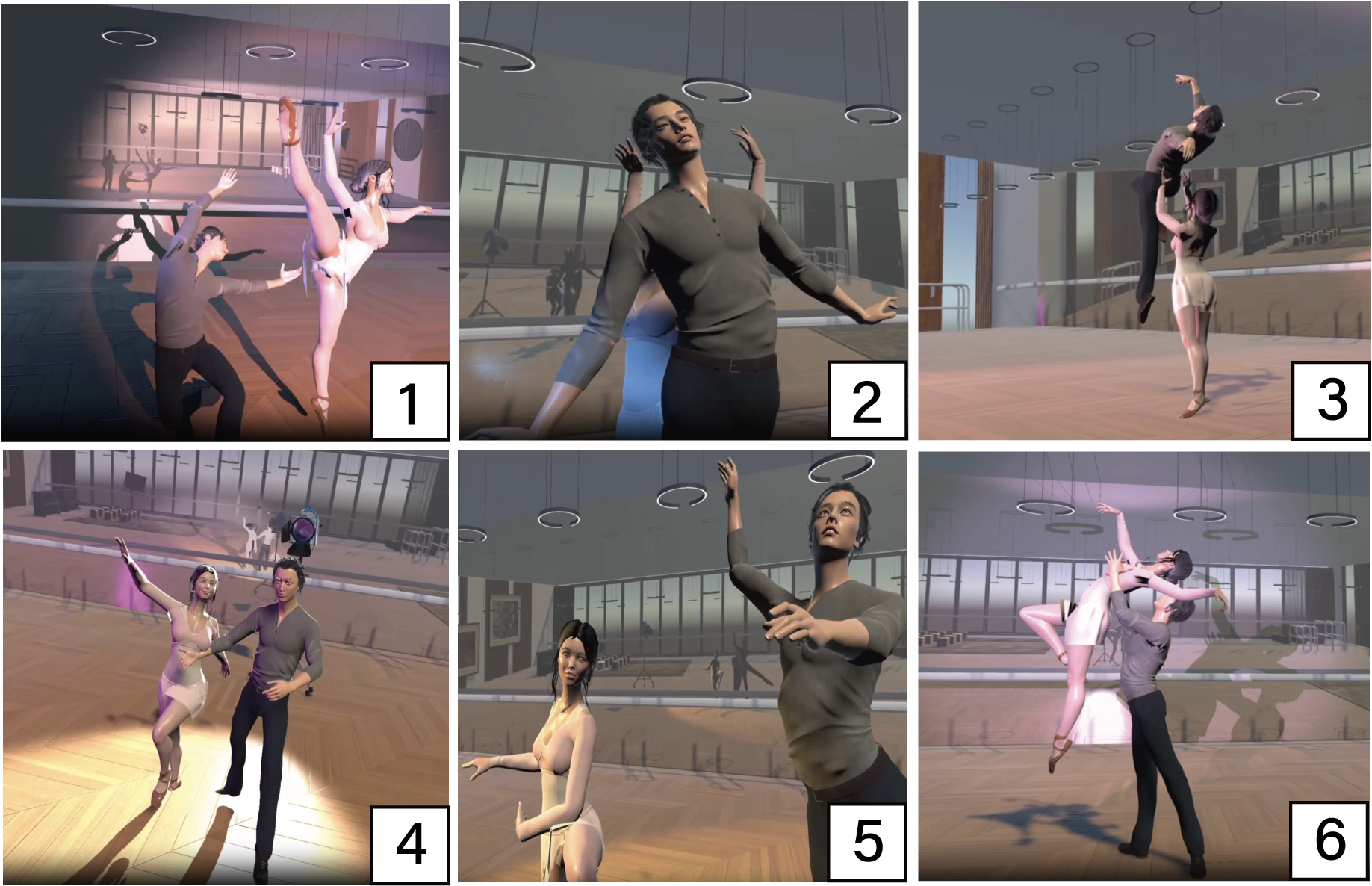}
    \caption{{Gaze exemplars: (1) A (M/M/M) wide shot with high-contrast rim light focused on the female dancer's body, male-gaze; (2) H (F/F/F) soft close-up on facial expression under even lighting, female-gaze; (3) D (M/F/F)  pastel mid-wide framing that lingers on hand interaction, female-gaze; (4) E (F/M/M) dynamic tracking shot with stark back-light highlighting motion, male-gaze; (5) B (M/M/F) centred mid-shot with soft fill and minimal body emphasis, neutral medium female-gaze; (6) G (F/F/M) sequence that opens with an intimate close-up and cuts to a kinetic wide lift, mixed male- and female-gaze conventions.}}
    \label{gaze}
\end{figure}

As shown in Figure \ref{expert}, two experts identified significant differences in the distribution of gaze categories across the eight groups. Group A, consisting of male users with male avatars and male actor-driven narratives (M/M/M), primarily produced works classified as ``male gaze,'' with only one piece labeled as ``female gaze'' or ``neutral.'' In contrast, Group B, featuring male users with male avatars but a female actor-driven narrative (M/M/F), predominantly exhibited the ``female gaze,'' making up more than half of its works. 

Groups C (M/F/M), D (M/F/F), and H (F/F/F) showed notably higher proportions of the ``female gaze'' but included a few works labeled as ``neutral'' or ``male gaze.'' Conversely, Groups E (F/M/M) and F (F/M/F) exhibited the opposite trend, with four out of six pieces in each judged as ``male gaze.'' Meanwhile, ``female gaze'' accounted for over half of Group G's (F/F/M) works. These variations reflect clear differences in how participants approached video creation and what aspects they chose to emphasize.

\begin{figure}[ht]
    \centering
    \includegraphics[width=8.5cm]{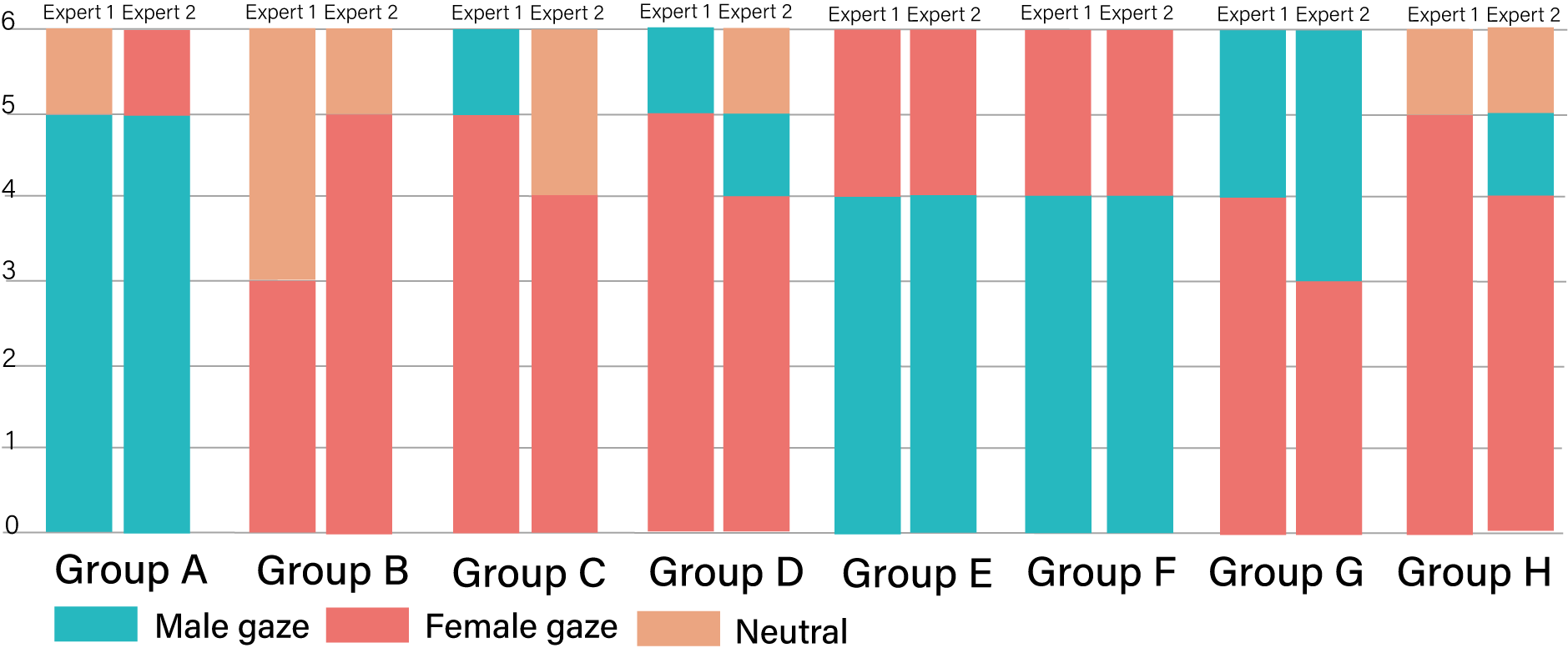}
    \caption{The judgments made by two experts on videos created by 48 participants across 8 groups, categorizing them into ``Male Gaze,'' ``Female Gaze,'' and ``Neutral.'' Each expert assigned a gaze category based on the presence of a significant gaze tendency in the videos. If the gaze tendency was unclear or insignificant, it was categorized as ``Neutral.'' The x-axis represents the 8 groups, while the y-axis shows the number of participants in each group.}
    \label{expert}
\end{figure}

As shown in Figure \ref{gaze} (1)-(6), this distribution pattern is closely tied to the participants' gender and the gender of the avatars they used. This evidence reinforces our hypotheses 2 and 3. Hypothesis 2 holds, especially when the avatar's gender matches the participant's; male participants tend to favor dynamic shots and high contrast to heighten motion and tension, whereas female participants prefer softer lighting and close-up shots to highlight emotional interactions. Results from the filmmaking tasks confirm these tendencies. On the other hand, Hypothesis 3 suggests that when the avatar's gender does not match the participant's, men who use female avatars focus more on emotional engagement and reduce emphasis on action or bodily details, while women who use male avatars lean toward visually striking, action-oriented shots. The experimental findings also support this conclusion.

\subsection{Observation and Semi-structured interviews}

{We conducted a qualitative analysis based on semi-structured interviews and direct observation. Two researchers independently reviewed all interview transcripts line by line, applying an inductive open coding strategy~\cite{corbin2015basics}. Recurring keywords were identified and iteratively grouped into broader themes. Any discrepancies in interpretation were discussed and resolved in weekly meetings with a senior researcher. Through this process, we found that a substantial portion of the interview data centered on participants' reflections on their avatars and their observations of the differences in the dancers' movements and how these differences influenced their filming decisions.}

\subsubsection{Observational Study}
Our observation indicates that male users tended to spend more time experimenting with, positioning, and adjusting the lighting. In contrast, female users were likely to focus on observing the dancers' movements and adjusting camera placement accordingly.


\subsubsection{Interview Findings}

\paragraph{{Perception of their own avatar.}} {Our interviews revealed that nearly all participants (N=46) observed their own avatars in the mirror.} Users who were assigned an avatar matching their gender tended to ignore their avatar, often describing it as \textit{``nothing special''(Group A (M/M/M))} or \textit{``I didn't really pay attention to it.''(Group G (F/F/M))} {In contrast, users with avatars of a different gender exhibited greater awareness of their avatars, often commenting that it felt different from real life. For instance, one participant said, ``I couldn't help but laugh when I saw my avatar in the mirror,'' noting that realizing the female avatar made him feel a clear disconnect from his real-life identity.} This avatar difference subtly influenced users' actions and behavior. One participant shared:

\textit{``Since I'm using a female avatar, I tend to perform in a more feminine way, for example, my camera movements align more with the direction of the female character, adjusting with the dancer's gaze.'' (Group D (M/F/F))}

\paragraph{{Perception of gender dynamics in the dancers’ movements.}} Participants {(N=41)} also noted apparent differences in the movements of the two dancers (actor-driven narrative). For example, some participants said: \textit{``I noticed the female dancer was lifting the male, which is different from real life,''(Group F (F/M/F))} and \textit{``The male dancer had a wider range of movement.''(Group E (F/M/M))} These differences influenced how users framed their shots. One participant shared, \textit{``Even though the male had more movement, I still focused on the female who was standing still,''(Group E (F/M/M))} and another explained, \textit{``Because the female was the one being lifted, my attention naturally went to her.''(Group G (F/F/M))} Others sought visual balance, keeping both dancers in frame despite their contrasting actions.

\subsubsection{{Operational definition of male vs.\ female gaze}} {Following feminist film theory and prior VR-gaze work \cite{cooper2000chick,dirse2013gender,oliver2017male}, Building on coding guidelines validated by expert raters, we operationalized three diagnostic cues for each gaze type. Male gaze was assigned when a shot exhibited at least two of the following: (i) emphasis on body or action (e.g., tracking a lift or lingering on torso/legs); (ii) high-contrast, directional lighting that foregrounds form; (iii) kinetic camera work (wide, dynamic moves). Female gaze was assigned when at least two of these were present: (i) focus on face, hands, or subtle interaction; (ii) soft, even lighting that foregrounds emotion; (iii) stable or intimate framing (close-up or medium). Shots meeting only one cue from each list were coded as neutral.}

\subsection{{Video Content Analysis}}
{Two researchers independently coded every 45-s participant clip frame-by-frame along four dimensions: (i) framing style (close-up, medium, wide), (ii) primary subject focus (male or female dancer), (iii) lighting (soft vs.\ high-contrast), and (iv) camera-movement emphasis, achieving substantial inter-coder agreement (Cohen's $\kappa = .78$).} {Artefacts paralleled expert ratings (Section \ref{Film Production Testing Task Results}; Fig.~\ref{gaze} \& \ref{expert}).} {{In Group A (M/M/M),} five of six clips relied on wide, high‑contrast takes that lingered on the female dancer's body or the lift sequence—classic ``male‑gaze'' compositions, fully consistent with their expert labels.} {{In Group H (F/F/F),} four clips favoured intimate close‑ups of facial expressions and hand gestures, coupled with soft key lighting, matching the ``female‑gaze'' categorisation.} {{As for cross‑gender embodiment (Groups C, D, E, F),} footage often shifted toward the \emph{avatar's} gendered style, when user and avatar gender diverged. For instance, \emph{Group D} (M/F/F) produced four clips with prolonged close-ups and pastel lighting, an unmistakable female gaze, whereas \emph{Group E} (F/M/M) emphasised dynamic tracking shots and stark contrast typical of a male gaze.} {{Narrative override (Groups B, G).} Even with gender‑matched avatars, a narrative led by the opposite gender modulated shot choice. \emph{Group B} (M/M/F) produced predominantly female‑gaze footage, while \emph{Group G} (F/F/M) balanced both perspectives, often beginning with an emotional close‑up before cutting to a wide action shot when the male dancer initiated the lift. For example, Figure \ref{gaze} (4), \emph{Group E} (F/M/M) begins with an overhead wide shot, moves to a static mid-shot of the dancer, and ends with a descending tilt that lands on the male performer’s final sprint. High-contrast rim lighting and dynamic framing foreground the body and movement, exemplifying a classic male-gaze aesthetic adopted by a female participant operating a male avatar.}

\section{Discussion}

\paragraph{VR's Influence on Gender Identity and Creativity}

The integration of gendered avatars and narrative perspectives in VR-based education influences not only the participants' creative decisions but also their gender perceptions and self-identification \cite{wu2024examining,bolt2021effects}. {Contrary to conventional expectations, participants using gender-incongruent avatars reported a stronger sense of presence and greater bodily awareness. Many described the experience as feeling markedly ``different from real life'' when seeing a virtual body that did not match their own, which led them to pay closer attention to their movements and framing choices. In contrast, gender-congruent conditions prompted less self-focus but revealed clearer gendered visual styles. Video analysis showed that female participants using female avatars tended to adopt soft lighting, close-ups, and emotionally expressive compositions, while male participants with male avatars favored high-contrast, wide-angle shots emphasizing motion. These tendencies were categorized by experts as ``female gaze'' and ``male gaze,'' suggesting that traditional visual habits persist even in immersive environments. Some participants reflected in interviews that this altered perception of body and role prompted them to consider ``who is narrating the story'' and the rationale for the camera's attention on certain characters. This awareness indicates that VR not only serves as a technical training tool but also fosters critical reflection on gendered perspectives \cite{smelik2007feminist,hollinger2012feminist,kors2016breathtaking}. By actively manipulating cameras and avatars, students are given a chance to identify and challenge their own narrative biases, i.e., developing more inclusive and critically engaged storytelling practices.}

\paragraph{The Proteus Effect in VR}
{The \textit{Proteus Effect} predicts that users' behaviour aligns with their avatar's traits \cite{szolin2023exploring,fox2013embodiment}, and our study extends this principle to VR film-making. Participants assigned gender-incongruent avatars tended to report slightly greater control and presence than those with gender-matched avatars and appeared to adjust their shooting style to align more closely with the avatar’s gender. Male students operating female avatars favoured ``female-gaze'' techniques, such as soft lighting and emotion-centred close-ups, whereas female students behind male avatars produced ``male-gaze'' footage with wide, high-contrast, action-oriented shots. Content analysis confirmed these cross-over patterns, directly supporting Hypothesis 3. A significant three-way interaction (user$\times$avatar$\times$narrative gender) emerged on the SEQ Awareness dimension, F(1, 40)= 4.28, p = 0.045. However, its directional pattern did not match our Hypothesis 4 prediction (that full gender alignment would yield the strongest effects) and did not replicate on other measures, so overall support for Hypothesis 4 remains limited. These patterns suggest that the \textit{Proteus Effect} may extend beyond surface-level behavior, potentially influencing participants’ artistic strategies. Pedagogically, deliberately assigning avatars that challenge learners' real-world identities can prompt exploration of unfamiliar visual styles, helping dismantle ingrained gender biases and fostering more flexible, empathetic creators \cite{crone2022interview,wu2024examining}. By offering diverse avatar options and story contexts, VR education can capitalise on this effect to broaden students' creative horizons and foster inclusive storytelling.}

\paragraph{{Educational Contribution to Film Pedagogy}}
{Although VR film-production drills are often viewed as purely technical \cite{wei2023feeling}, our study shows they also teach narrative perspective. Cross-gender embodiment prompted students to film from the opposite gender's gaze. Interview remarks such as ``I noticed who was telling the story...'' and our video analysis confirm that this role-play encouraged critical self-reflection on bias. Learners sometimes blended both styles, most notably in Group G, where a female user with a female avatar filmed a male-led narrative that opened with an intimate close-up and then shifted to a dynamic wide shot, producing a hybrid gaze rarely achieved in traditional classes. Thus, \textit{Film DanceStudio} is more than a camera-and-lighting simulator; it serves as a ``perspective playground.'' By allowing students to swap their avatars' gender and narrative focus, instructors can assign the same scene with different gender configurations and then guide a reflective discussion or peer critique. This systematic manipulation of viewpoint helps students recognizec and dismantle unconscious biases, fostering more versatile filmmakers. Although our case study centers on cinematography, the same avatar–narrative manipulations could be applied to theatre blocking, dance choreography, or interactive media design courses, where viewpoint and embodiment likewise shape creative choices.}

\paragraph{Avatar Design Enhances Engagement}
{Across the IPQ, learners using opposite-gender avatars felt markedly more ``present'' and scored higher on SEQ Awareness, indicating closer monitoring of their virtual bodies. This pattern contradicts Hypothesis 1, which predicted higher presence for gender-congruent avatars, indicating that a gender mismatch can create a productive form of discomfort that deepens cognitive engagement. Interviewees echoed this finding, noting they ``watched every movement'' when their avatar's gender differed from their own.} {This identity-driven engagement had a clear influence on cinematic decisions. We observed that avatar gender subtly steered how participants framed and lit their shots. Learners tended to align their filmmaking style with the perspective of their avatar's gender. Notably, when participants stepped into an avatar of the opposite gender, they often adopted the visual style associated with that avatar's gender. Participants even reflected on these shifts in post-experiment interviews, acknowledging that the avatar's identity made them rethink who ``tells the story'' in a scene and adjust their focus accordingly. In essence, the avatar's gender acted as a lens through which learners approached storytelling, influencing everything from shot composition to lighting tone.} {Our results, albeit limited by small participant numbers, offer clues to instructional designs, as follows. Using gender-challenging avatars increased presence and led students to make more deliberate camera choices. Allowing learners to select or switch among avatars of different genders helps them question default viewpoints and engage more fully with the task. Therefore, VR platforms should offer flexible avatar sets, allowing students to either affirm their own identity or experiment with another, i.e., both approaches have been shown to boost engagement.}

\paragraph{{Implications for Gender-Inclusive VR Pedagogy}}
{Our findings offer preliminary insights that may inform future VR studies, especially those exploring creative domains like film-production education.} Traditionally, VR educational research has primarily focused on the benefits of immersive, realistic avatars in enhancing learning outcomes \cite{schmidt2024frankenstein,you2024sense}. {We show that learner engagement and creative output also depend on the three-way interplay among avatar gender, participant gender, and narrative gender. When these elements align or clash, they alter cognitive focus, emotional connection, and cinematographic choices. 
Future studies should therefore move beyond ``one-size-fits-all'' avatars and investigate how subtle variations in avatar identity influence both technical and artistic performance \cite{ratan2020avatar,kim2024self}. In addition, VR environments that embed diverse gender narratives can prompt learners to question conventional portrayals and think critically about representation \cite{wu2024examining}.} This opens a promising direction for research, particularly in exploring how VR can be used to challenge traditional gender norms and encourage learners to reflect on the social and cultural impact of gender narratives in film \cite{myisha2023decoding}. Future research should investigate how VR can serve as a platform for deconstructing gender roles and stereotypes, offering students the opportunity to experiment with non-traditional gender perspectives in media production.

\paragraph{{Design Considerations for VR System}}
For VR developers, this study highlights several critical areas for improvement to create more engaging, meaningful, and inclusive educational environments. One of the most significant findings is the importance of seamlessly integrating narrative and avatar design choices within VR systems to ensure a more immersive and interactive learning experience. Developers should prioritize the ability to offer flexible avatar customization, allowing learners to select avatars that reflect diverse gender identities, ethnicities, and personal preferences \cite{do2024stepping,do2024cultural}; {Our analysis could potentially relates to two concrete design improvements for fostering richer and more inclusive learning: (1) allow learners to switch quickly among avatars spanning a spectrum of gender identities, and (2) provide each practice scene in both male-led and female-led narrative variants. When these cues were mixed (e.g., Group G), students blended male- and female-gaze conventions, indicating that tension between avatar identity and story viewpoint can spark deeper visual experimentation. Combining flexible avatar options with dual narrative paths allows a VR camera-and-lighting sandbox to teach technical skills while prompting learners to reflect on how gendered viewpoints shape cinematic language.}

\paragraph{Limitations and Future Work}

This study reveals how participant, avatar, and narrative gender shape creative choices in VR filmmaking, though several limitations affect the findings' interpretation and generalizability. First, the relatively small sample size {($N = 48$; $n = 6$ per cell) limits statistical power and thus constrains the generalizability of our findings. A post-hoc power analysis for the observed interaction (partial $\eta^{2} = 0.09$, corresponding to $ 0.31$; $\alpha = 0.05$; 8 groups) yields\textit{power} $= 0.25$, below the conventional 0.80 benchmark.} The sample was predominantly composed of students from a single academic institution, which reduces the external validity of the findings when considering more diverse learner populations. Additionally, the lack of diversity in terms of prior experience, demographic characteristics, and team gender composition further limits the applicability of these findings across broader groups. Post-hoc power analysis suggests that to detect larger effects (f = 0.35) with adequate statistical power (80\%), a sample size of approximately 125 participants would be necessary. Future research should therefore aim to incorporate larger and more heterogeneous samples, accounting for variations in educational background and cultural context. This would enable a more thorough examination of the trends and a deeper understanding of the interactions between avatar characteristics, narrative roles, and user behavior. Second, the pre-defined avatar designs used in this study, while aligned with participants' ethnicities and skin tones, may not fully capture the broad spectrum of avatar customization options that reflect real-world diversity and user preferences. These avatars were not personalized to individual users, which may affect the ecological validity of the study's findings \cite{do2024cultural}. Furthermore, {we used binary-gender avatars to establish a clear male/female gaze baseline; adding androgynous avatars, shown in earlier VR studies to shift embodiment and bias \cite{freeman2021body,zhang2024gender}, would introduce another identity factor and complicate attribution. With this baseline set, future work can include non-binary avatars to investigate fluid or hybrid gaze strategies,} and expand our understanding of avatar design's role in shaping VR interactions and creative outcomes. The study also focused exclusively on the effects of avatar gender on individual gaze patterns. Subsequent research could examine how the gendered avatars of multiple users influence gaze dynamics in collaborative settings, shedding light on how gendered perspectives are negotiated in group-based creative tasks. In addition, reliance on self-reported measures, such as IPQ and SEQ, introduces the potential for response biases, including social desirability bias or inaccurate recall. While these subjective measures provide valuable insights into participants' perceptions, they are inherently limited by the participants' self-awareness and the potential distortion of responses. This study focused on avatar gender, but other cues—like body type, facial expressions, and voice—may also shape user behavior \cite{kim2024self,banakou2013illusory}. Future work should explore how these factors, individually or combined, influence interaction and creativity in VR.

\section{Conclusion}
This study shows that avatar and narrative gender jointly shape engagement in VR film production. Gender mismatches between users and avatars affect immersion and influence creative choices: female users with cisgender avatars favored emotional shots, while males preferred action-driven ones. Notably, gender-incongruent avatars prompted more thoughtful, creative decisions, suggesting deeper cognitive engagement. VR thus offers a unique space to challenge gender norms, and future systems should allow users to explore diverse avatar and narrative genders to foster critical reflection on representation.

\bibliographystyle{abbrv-doi-hyperref}

\bibliography{ISMAR}
\end{document}